\documentclass[aps,amsmath,amssymb,amsthm,longbibliography,prl]{revtex4-2}

\usepackage{graphicx}% Include figure files
\usepackage{dcolumn}% Align table columns on decimal point
\usepackage{bm}% bold math
\usepackage{color}
\usepackage{soul}
\soulregister\ref7
\soulregister\cite7
\usepackage{marvosym}
\usepackage{makecell}
\usepackage{tabularx}

\usepackage[mathlines]{lineno}% Enable numbering of text and display math
\begin{document}

%\preprint{APS/123-QED}

\title{Quantum walk with coherent multiple translations induces fast quantum gate operations}

\author{Yixiang Zhang$^{1,*}$, Xin Qiao$^{1,2,*,\dagger}$, Luojia Wang$^{1}$, Yanyan He$^{1}$, Zhaohui Dong$^{1}$, Xianfeng Chen$^{1,3,4}$, Luqi Yuan$^{1,\ddagger}$}

\affiliation{
$^1$State Key Laboratory of Photonics and Communications, School of Physics and Astronomy, Shanghai Jiao Tong University, Shanghai 200240, China\\
$^2$College of Physics and Electronics Engineering, Northwest Normal University, Lanzhou 730070, China\\
$^3$Collaborative Innovation Center of Light Manipulations and Applications, Shandong Normal University, Jinan 250358, China\\
$^4$Shanghai Research Center for Quantum Sciences, Shanghai 201315, China\\
$^*$These authors contributed equally.\\
$^{\dagger}$Corresponding author: qiaox@nwnu.edu.cn\\
$^{\ddagger}$Corresponding author: yuanluqi@sjtu.edu.cn
}

%\begin{abstract}
%Quantum walks with one-dimensional translational symmetry are important for quantum algorithms, where the speed-up of the diffusion speed can be reached if long-range couplings are added. Our work studies a scheme of a ring under the strong resonant modulation that can support discrete-time quantum walk including coherent multiple long-range translations in a coherent way along synthetic frequency dimension. These multiple translation paths are added in a coherent way, which makes the walker evolve under the topological band. Therein, not only the fast diffusion speed is expected, but more importantly, we find that single quantum gate operations can be performed in the quasi-momentum space. In particular, we show the arbitrary single-qubit state preparation and an example of CNOT two-qubit gate with only one time step, dramarically increasing quantum algorithms. Our study uses a single ring to provide fast quantum gate operations based on coherent multiple path quantum walk, which may provide unique designs for efficient quantum operations on photonic chips.
%\end{abstract}

\maketitle

\section{Abstract}
Quantum walks with one-dimensional translational symmetry are important for quantum algorithms, where the speed-up of the diffusion speed can be reached if long-range couplings are added. Our work studies a scheme of a ring under the strong resonant modulation that can support discrete-time quantum walk including coherent multiple long-range translations in a natural way along synthetic frequency dimension. These multiple translation paths are added in a coherent way, which makes the walker evolve under the topological band. Therein, not only the fast diffusion speed is expected, but more importantly, we find that single quantum gate operations can be performed in the quasi-momentum space. In particular, we show the arbitrary single-qubit state preparation and an example of CNOT two-qubit gate with only one time step, dramarically increasing quantum algorithms. Our study uses a single ring to provide fast quantum gate operations based on coherent multiple path quantum walk, which may provide unique designs for efficient quantum operations on photonic chips.

\section{Introduction}
The concept of quantum walk \cite{PhysRevA.48.1687,PhysRevA.58.915,10.1080/00107151031000110776,Venegas-Andraca2012} has been developed for various subjects including the quantum algorithm design \cite{10.1080/00107151031000110776,PhysRevA.67.052307,10.1142/S0219749903000383,doi:10.1137/S0097539705447311,Chawla2020, doi:10.1137/1.9781611977073.109}, quantum dynamics simulation \cite{PhysRevLett.94.100602,PhysRevA.83.022320,Huerta_Alderete2020}, and topological phases exploration  \cite{PhysRevA.82.033429,PhysRevB.84.195139,Cardano2016,PhysRevA.103.012201,PhysRevA.97.052117,Mittal_2021,PhysRevA.98.013835,PhysRevA.107.032201}. Due to quantum interference, quantum walk displays different behaviors than classical random walk \cite{PhysRevA.48.1687,10.1145/380752.380758,10.1145/380752.380757,PhysRevA.58.915}. One of the key features is that the quantum walk spreads quadratically faster than its classical counterparts \cite{10.1145/380752.380758,10.1145/380752.380757}, which makes it possible for realizing a version of the Grover's search algorithm \cite{grover1996fastquantummechanicalalgorithm,PhysRevLett.79.325} with a square-root reduction in the execution time compared to classical algorithms \cite{ambainis2004coinsmakequantumwalks}. Therefore, it has been well recognized that the quantum speed-up of quantum walk is crucial for quantum computing. Due to the interference nature behind \cite{Fenwick24}, photonic systems provide important platforms for performing quantum walks, including experiments in bulk optics \cite{PhysRevLett.104.153602,RN69,Nejadsattari19,DErrico20,DiColandrea:23}, fiber loops \cite{PhysRevLett.104.050502,doi:10.1126/science.1218448,PhysRevResearch.3.023183,PhysRevA.105.042210}, fiber cavities \cite{Boutari_2016}, and integrated photonics \cite{PhysRevLett.108.010502,Gräfe_2016}.

It is worth noting that the efficiency of quantum walk in photonic lattices may be further sped up algorithmically if the long-range coupling between lattice sites is added. Especially for the universal quantum gate designs using quantum walks for qubits on the graph \cite{PhysRevLett.102.180501,PhysRevA.81.042330,10.1126/science.1229957,Singh2021}, it usually requires complex geometric structures in photonics \cite{Souza2022,Mandal2023}. Hence efficient simplification of a photonic configuration in realizing quantum walk including long-range couplings is essential for realizing quantum computing with desired quantum-gate functionality in integrated photonics.

The synthetic frequency dimension built in rings under dynamic modulations may provide a solution as it can provide artificial lattice model with long-range couplings and may be implemented with recent state-of-art technology on photonic chips \cite{Hu20,RN75,RN76,RN77,RN78}. Although various photonic simulations have been demonstrated in synthetic frequency lattice models \cite{Yuan:16,RN83,RN73}, current researches mainly focus on the weak modulation limit and hence the models obey the continuous-time Schrödinger equation \cite{RN83,RN73,RN84,PhysRevLett.130.083601,RN70}. Therefore, it is of fundamental importance to study the synthetic frequency lattice under the strong modulation regime and the resulting discrete-time quantum walk algorithm therein.

In this paper, we theoretically present the one-dimensional discrete-time quantum walk (1D DTQW) in the synthetic frequency dimension built by a photonic ring under the strong electro-optic modulation (EOM). It is then noted that translation operators with coherent multiple long-range transition effects, different from the previous works with additional long-range couplings \cite{RN75,RN77,RN78}, are naturally introduced into the quantum walk, which leads the topological band and enhancement of the walker's diffusion speed along the frequency dimension. Since the variation of the field is large for roundtrip time it circulating inside the ring under the strong modulation regime, we develop the discrete-time quantum walk for our model, instead of the continuous-time Schrödinger  equation in previous works with weak modulation limit \cite{RN83,RN73,RN84}. Moreover, by tuning parameters of EOM, the band in the quasi-momentum space ensures us to construct single quantum gate operations and prepare single-qubit/two-qubit states in the fast manner. Our work reveals the coherent way of quantum interference among multiple long-range translations in the quantum walk with synthetic frequency lattice, which may greatly improve the speed in quantum algorithms.

\section{Results}

\subsection{Quantum walk}

\begin{figure}[htbp]
\centering
\includegraphics[width=8.5cm]{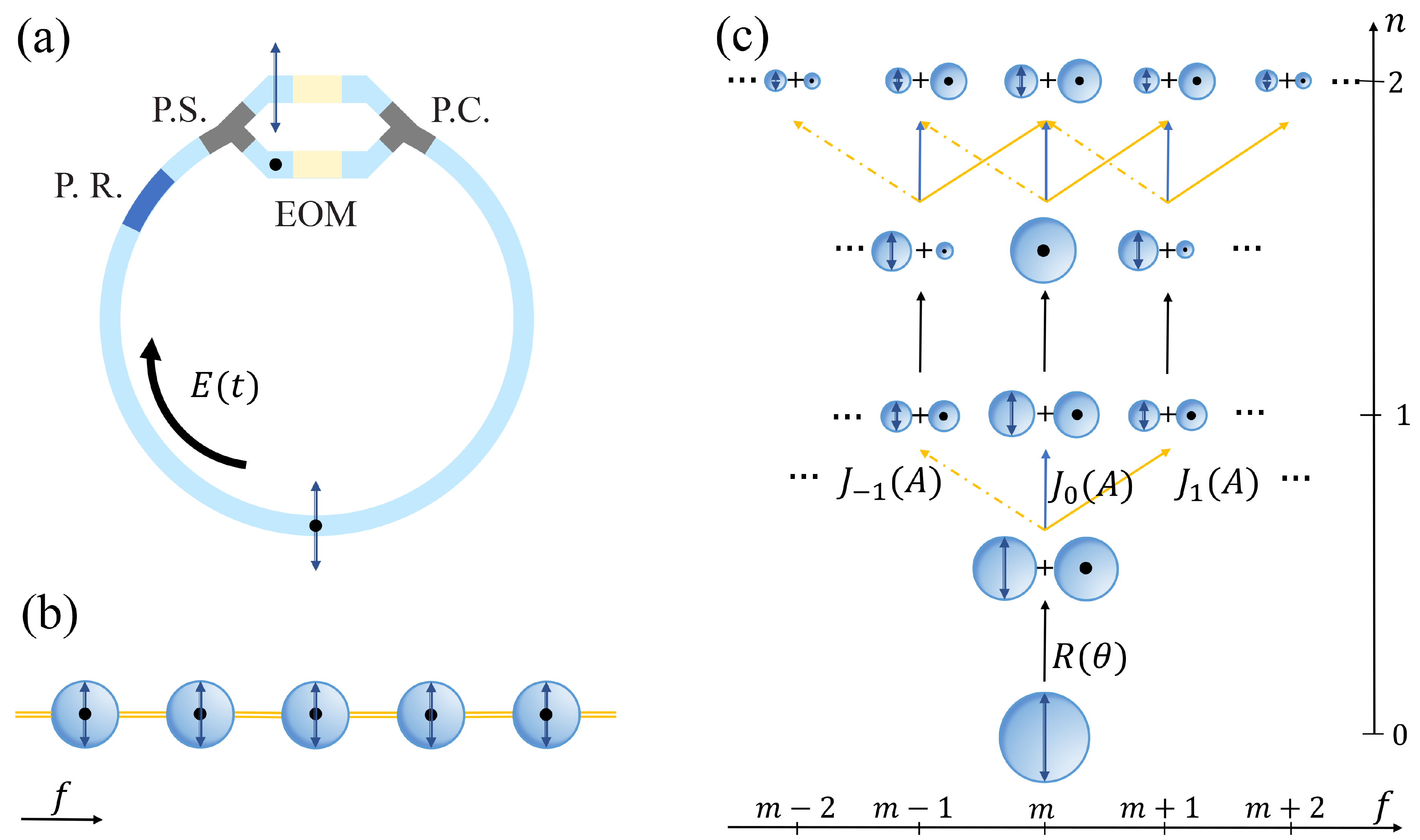}% Here is how to import EPS art
\caption{\label{fig:fig1} (a) Ring resonator with a polarization rotator (P.R.), a polarization splitter (P.S.), two EOMs, and a polarization combiner (P.C.). The dots and black arrows represent $H$- and $V$-polarization respectively. (b) One-dimensional synthetic spin lattice in the frequency dimension. (c) Schematic of 1D DTQW in the synthetic frequency dimension. The area of each circle represents the magnitude of the polarization-dependent amplitude $\Psi_H(t)$ and $\Psi_V(t)$.}
\end{figure}

We consider a ring resonator, composed by the photonic waveguide that can support two polarizations of light, as shown in Fig. \ref{fig:fig1}(a), where the two orthogonal polarization modes of light (labelled by $H$ and $V$) form the pseudo-spin state basis for the electromagnetic waves. A polarization splitter is added into the ring to split light field to $H$- and $V$-polarization components respectively and then each component is propagating through the corresponding polarization-maintaining waveguide branches [see Fig. \ref{fig:fig1}(a)]. Afterwards, field components from two branches combine into the main section of the ring via a polarization combiner. The similar design has been previously proposed and recently experimentally used to study non-Abelian gauge field \cite{PhysRevLett.130.083601,RN70,RN72}.

The ring supports frequency resonant modes with the free spectral range $\Omega$ for both polarization components if the group velocity dispersion of the waveguide is ignored. Each pair of modes at the same frequency holds an effective spin site. By applying dynamic modulation in each branch, one can connect the spin state at each site in a desired way and hence construct the synthetic spin lattice in the frequency dimension of light [see Fig. 1(b)]. A polarization rotator is added to rotate the polarization of the field inside the ring, i.e.,  to rotate the spin state in the synthetic lattice. 

We here provide the mathematical description of the field dynamics in our designed ring. We first expend the electric field inside the ring as
\begin{equation}\label{eq1}
\Psi(t)=\sum_m a_m (t) e^{i\omega_m t}.
\end{equation}
Here $\Psi\equiv[\Psi_H, \Psi_V]^T$ is the polarization-dependent amplitude of the electric field and $a_m\equiv[a_{m,H}, a_{m,V}]^T$ is the polarization-dependent amplitude of the $m$th resonant mode at the resonant frequency $\omega_m \equiv \omega_0 + m\Omega$. $\omega_0$ is a reference resonant frequency in the ring, which can usually be set as zero for simplicity \cite{RN73}. One can replace the time $t$ by $t=nT_R + t_f$, where $T_R=2\pi/\Omega$, $n$ is a non-negative integer representing the number of round-trips, and $t_f\in[-T_R/2,T_R/2)$ is the travel time of light in each round-trip. The variation of the field after it finishes each round-trip can be described by
\begin{equation}\label{eq2}
\begin{aligned}
\Psi(t+T_R)=&\sum_m a_m(t+T_R)e^{i\omega_m t}\\
=&D(t)R(\theta)\sum_m a_m(t)e^{i\omega_m t},
\end{aligned}
\end{equation}
where $R(\theta)$ and $D(t)$ takes the form
\begin{equation}\label{eq3}
R(\theta)\equiv e^{-i\theta \sigma_y/2}=
\left(
\begin{array}{cc}
\cos\frac{\theta}{2} & -\sin\frac{\theta}{2} \\
\sin\frac{\theta}{2} & \cos\frac{\theta}{2} \\
\end{array}\right)\;,
\end{equation}
\begin{equation}\label{eq4}
D(t)\equiv
\left(
\begin{array}{cc}
e^{i\mathit{\Gamma}\cos(\Omega t+\phi_H)} & 0 \\
0 & e^{i\mathit{\Gamma}\cos(\Omega t+\phi_V)} \\
\end{array}\right)\;,
\end{equation}
describing the polarization rotation and polarization-dependent modulations in Fig. \ref{fig:fig1}(a), respectively. Here $\theta$ is the polarization rotation. Two EOMs are under the resonant phase modulation at the modulation strength $\mathit{\Gamma}$ with different modulation phases $\phi_H$ and $\phi_V$. 

This system can be viewed as a quantum walk process described by the unitary step operator. In the constructed synthetic frequency lattice, the basis vector is $|m, p\rangle$, where $m$ is the coordinate of the walker in the synthetic frequency dimension, and $p$ is the spin state (the coin state \cite{10.1080/00107151031000110776}) based on $|H\rangle$ and $|V\rangle$. Therefore, by placing Eq. (\ref{eq1}) into Eq. (\ref{eq2}) and keeping terms at the same $\omega_m$, the unitary step operator $U$ after each roundtrip reads $U=TR(\theta)$, where $T$ denotes a polarization-dependent translation operator in synthetic space from Eq. (\ref{eq4}):
\begin{equation}\label{eq10}
\begin{aligned}
T=&\sum_{m,l=-\infty}^{\infty}i^lJ_l(\mathit{\Gamma})e^{il\phi_H}|m+l\rangle\langle m|\otimes|H\rangle\langle H|\\
&+\sum_{m,l=-\infty}^{\infty}i^lJ_l(\mathit{\Gamma})e^{il\phi_V}|m+l\rangle\langle m|\otimes|V\rangle\langle V|.
\end{aligned}
\end{equation}
$J_l(\mathit{\Gamma})$ is the $l$-th-order Bessel function. The operator $U$ reflects the physics of the original periodically driven system in Eq. (\ref{eq2}). The dynamics of such quantum walk process can be understood by the schematic in Fig. \ref{fig:fig1}(c), where the field at the $m$-th resonant mode gets the rotation operation ($R$) and then undergoes the polarization-dependent translation operation ($T$) for a single roundtrip. Note this illustration in Fig. \ref{fig:fig1}(c) only shows the walker (the wave function of the field in the synthetic frequency lattice) transit to adjacent modes under the weak modulation strength $\mathit{\Gamma}$, which may derive to the continuous-time Schrödinger equation in previous works \cite{RN83,RN73,RN84}. Nevertheless, once $\mathit{\Gamma}$ is large, the derivation of the continuous-time Schrödinger equation is no longer valid (see the supplementary note \uppercase\expandafter{\romannumeral1} for details). The dynamics of our model is discrete with long-range transitions at different coupling strengths $J_l(\mathit{\Gamma})$ on the translation operation include in a coherent way.

To understand the quantum walk in the synthetic frequency lattice, we transform $T$ into quasi-momentum space which is reciprocal to the frequency and in the unit of time, and obtain
\begin{equation}\label{eq11}
\begin{aligned}
&U_k=\int_{BZ}dk\\
&\times\left(
\begin{array}{cc}
e^{i\mathit{\Gamma}\cos(k+\phi_H)}\cos\frac{\theta}{2} & -e^{i\mathit{\Gamma}\cos(k+\phi_H)}\sin\frac{\theta}{2} \\
e^{i\mathit{\Gamma}\cos(k+\phi_V)}\sin\frac{\theta}{2}& e^{i\mathit{\Gamma}\cos(k+\phi_V)}\cos\frac{\theta}{2} \\
\end{array}\right)\;\otimes|k\rangle\langle k|,
\end{aligned}
\end{equation}
where $k$ is the quasimomentum and is exactly $t_f$ \cite{RN79}. By applying the Floquet band theory \cite{PhysRevA.15.1109} and using $U=e^{-iH_{\textup{eff}}T_R/\hbar}$, we obtain an effective Hamiltonian in quasimomentum space:
\begin{equation}\label{eq13}
\begin{aligned}
H_{\textup{eff}}=\frac{\Omega}{2\pi}\int_{BZ}dk[E(k)\bold{n}(k)\cdot\boldsymbol{\sigma}]\otimes|k\rangle\langle k|.
\end{aligned}
\end{equation}
Here $E(k)$ is the quasienergy satisfies
\begin{equation}\label{eq14}
\begin{aligned}
\cos E(k)=&\pm\cos\Big[\frac{\mathit{\Gamma}}{2}(\alpha+\beta)\Big]\cos\frac{\theta}{2}\cos\Big[\frac{\mathit{\Gamma}}{2}(\alpha-\beta)\Big]\\
&\mp\sin\Big[\frac{\mathit{\Gamma}}{2}(\alpha+\beta)\Big]\sin(k,\phi_H,\phi_V),
\end{aligned}
\end{equation}
where $\alpha=\cos(k+\phi_H)$, $\beta=\cos(k+\phi_V)$ and
\begin{equation}\label{eq16}
\begin{aligned}
\sin(k,\phi_H,\phi_V)=\sqrt{1-\bigg\{\cos\Big[\frac{\mathit{\Gamma}}{2}(\alpha-\beta)\Big]\cos\frac{\theta}{2}\bigg\}^2}.
\end{aligned}
\end{equation}
$\boldsymbol{\sigma}=(\sigma_x,\sigma_y,\sigma_z)$ and $\bold{n}=(n_x,n_y,n_z)$ are the Pauli matrix vector and polarization eigenstate vector, respectively (see the supplementary note \uppercase\expandafter{\romannumeral2} for details). $\hbar=1$ for the simplicity.

\begin{figure}[t]
\centering
\includegraphics[width=9cm]{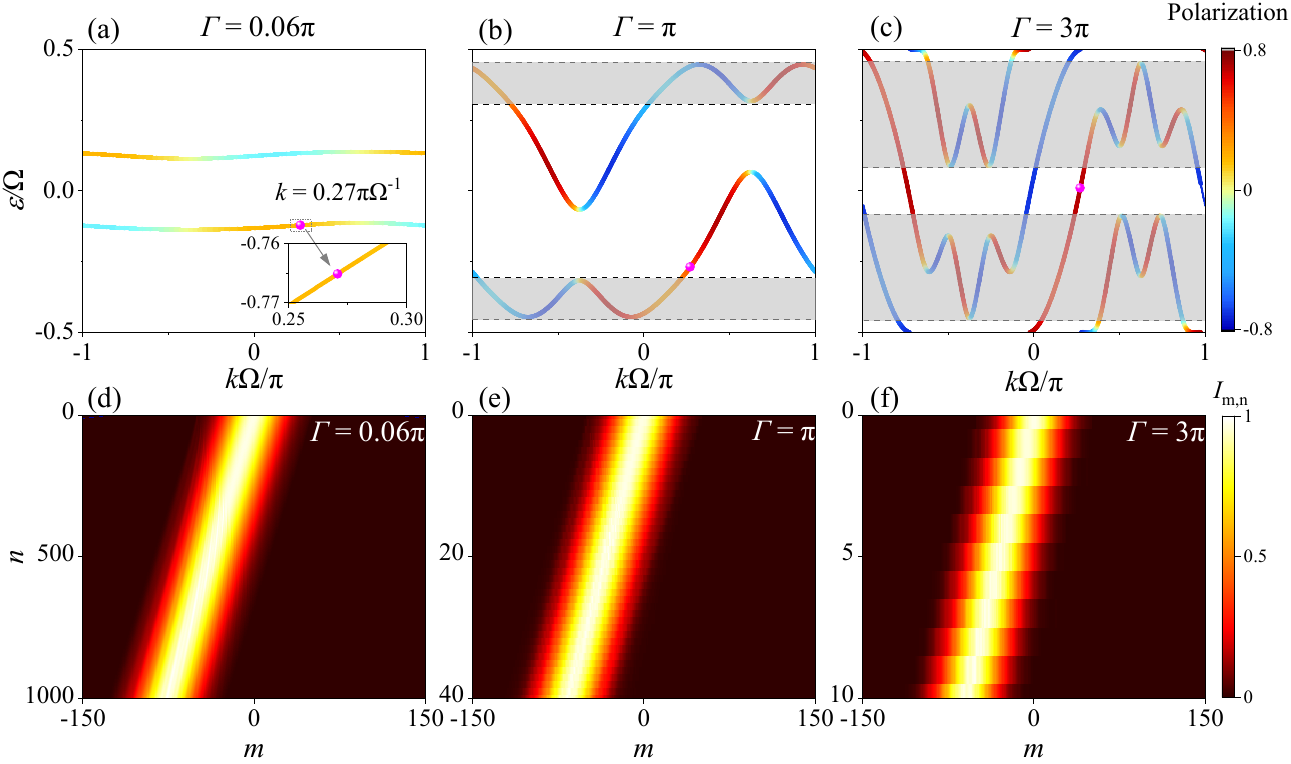}% Here is how to import EPS art
\caption{\label{fig:fig2} Quasienergy spectrum of the Hamiltonian $H_{\textup{eff}}$ with $(\theta,\phi_H,\phi_V)=(-\pi/2,0,3\pi/4)$ and different modulation intensities: (a) $\mathit{\Gamma}=0.06\pi$, (b) $\mathit{\Gamma}=\pi$, (c) $\mathit{\Gamma}=3\pi$. The colorbar specifies the projection of eigenstates on $H$- (or $V$-) polarization with the value up to 1 ($-1$). (d)-(f) Evolutions of eigenstates excited at quasi-momentum $k=0.27\pi \Omega^{-1}$ in simulations with different $\mathit{\Gamma}$ in simulations.}
\end{figure}

We plot the quasienergy spectra from $H_{\textup{eff}}$ with various choices of $\mathit{\Gamma}$ in Fig. \ref{fig:fig2}(a)-\ref{fig:fig2}(c). For Fig. \ref{fig:fig2}(a), we choose $\mathit{\Gamma}=0.06\pi$, i.e., the weak modulation, with parameters $\theta=-\pi/2$, $\phi_H=0$, and $\phi_V=3\pi/4$. One can see the typical topological edge modes associated to the quantum Hall ladder, where the spectrum exhibits one-way dispersion for the projection onto one polarization. Note the range of each band in the spectrum is $0.02\Omega$ due to the small $\mathit{\Gamma}$. Such range is greatly enlarged once $\mathit{\Gamma}$ becomes larger [see Figs. \ref{fig:fig2}(b) and \ref{fig:fig2}(c)]. In Fig. \ref{fig:fig2}(b) with $\mathit{\Gamma}=\pi$, the bands expand over the spectral first Brillouin zone (FBZ) $\varepsilon \in [-0.5,0.5]\Omega$, and they present more concentration on one polarization. Moreover, at a portion of $k$, the one-way edge modes on one band projected on the same polarization are paired with opposite dispersions for a given $\varepsilon$ [see the shadow regime in Fig. \ref{fig:fig2}(b)], which is fundamentally different from that in Fig. \ref{fig:fig2}(a), as in this case the excited edge mode may be scattered to the one at the opposite direction in principle. However, one can see that there are still part of edge modes that are free of such pairing. In addition, large $\mathit{\Gamma}$ results in greater steepness, which leads to the larger speed of the quantum walker evolution. Lastly, if $\mathit{\Gamma}$ is further enlarged, the bands outside of the spectral FBZ get into FBZ, for example as shown in Fig. \ref{fig:fig2}(c) with $\mathit{\Gamma}=3\pi$. The spread speed of the quantum walker gets further enhanced and the regime covering the similar pairing on one band is also enlarged. Here we give several additional notes. Firstly, the pairing is limited in edge modes on the same band. In principle, the two separate bands are protected by different topological invariants if bulks are introduced. Secondly, for one band in Fig. \ref{fig:fig2}(c), the edge mode projected onto $H$- (or $V$-) polarization supports three portions with negative dispersion and two portions with positive dispersion, and vice versa. Therefore, the pairing here can be multiple modes with opposite dispersions.

In simulations, we excite the walker using a Gaussian-shape wave packet $\bold{s}(m)=e^{-m^2/\Delta^2}\cdot e^{-ikm\Omega}\bold{e_s}$ in the synthetic lattice, where $\bold{e_s}=\big(\langle H|\bold{n}(k)\cdot\boldsymbol{\sigma}\rangle,\langle V|\bold{n}(k)\cdot\boldsymbol{\sigma}\rangle\big)^T$ carries the initial polarization information. We apply the unitary step operator $U$ to the initial state $|\bold{s}\rangle$ to simulate light circulating in the ring for one roundtrip. We choose corresponding $\bold{e_s}$ and $\Delta=25$, $k=0.27\pi \Omega^{-1}$ to excite the eigenstates labelled by pink dots in Figs. \ref{fig:fig2}(a)-2(c) and plot the wave function distribution $P(m,n)=\sum_{p}\big{|}\langle m,p|U^n|\bold{s}\rangle\big{|}^2$ after $n$-th step (roundtrip) in Fig. \ref{fig:fig2}(d)-2(f). One sees all evolutions exhibit unidirectional frequency conversion with similar patterns, but the one with larger $\mathit{\Gamma}$ shows significant faster evolution speed of the quantum walker in the synthetic frequency lattice.

\begin{figure}[hbpt]
\centering
\includegraphics[width=9cm]{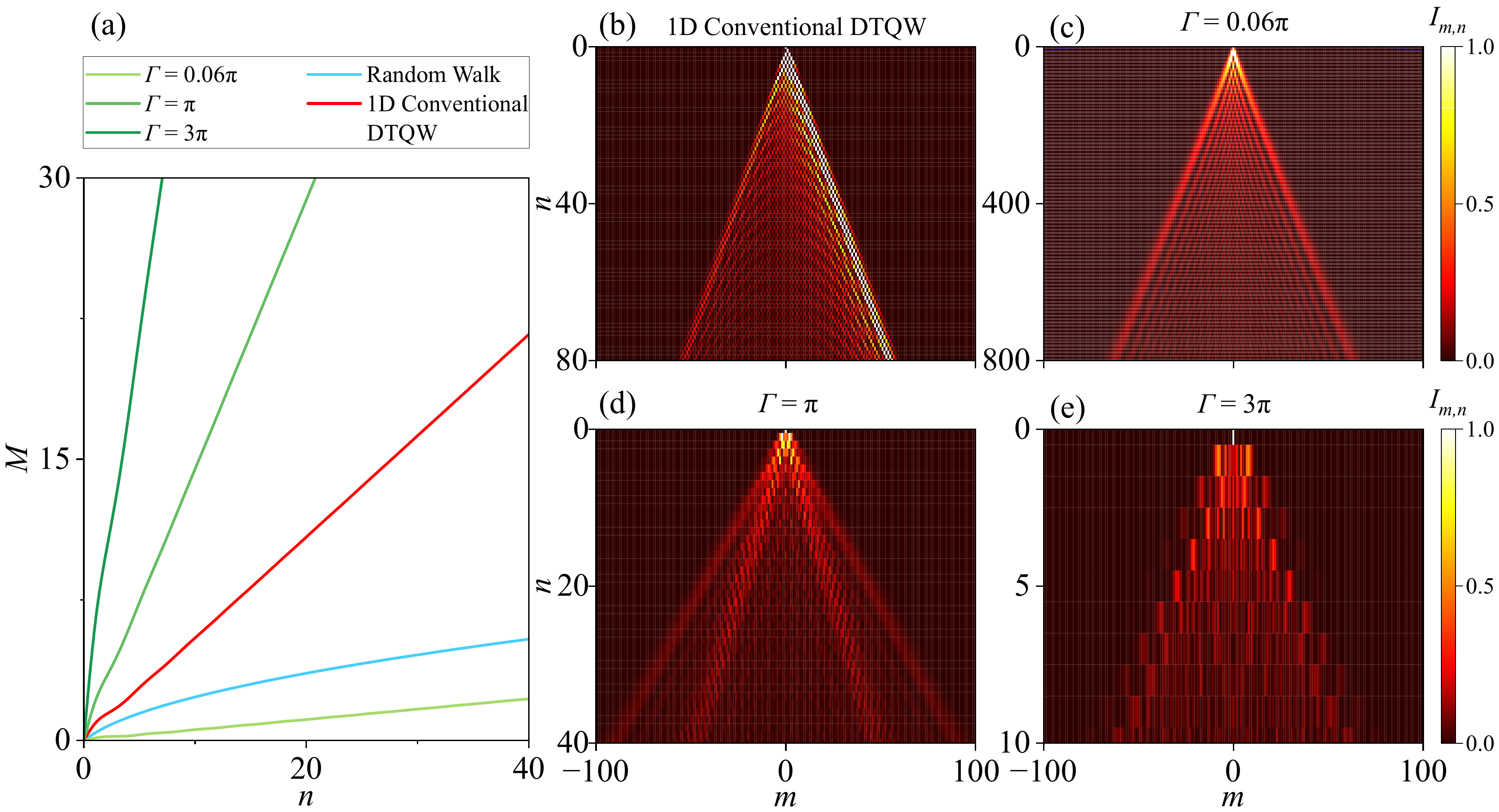}
\caption{\label{fig:fig3} (a) Diffusion distances calculated in the classical random walk (blue), 1D conventional DTQW (red), and 1D DTQW in the frequency dimension with three different $\mathit{\Gamma}$ (light green, green, and dark green, respectively) versus step (roundtrip) number $n$. Evolutions of (b) 1D conventional DTQW, and (c)-(e) 1D DTQW in the synthetic frequency dimension with (c) $\mathit{\Gamma}=0.06\pi$, (d) $\mathit{\Gamma}=\pi$, (e) $\mathit{\Gamma}=3\pi$ in simulations.}
\end{figure}

To further compare the speed of the wave-function diffusion between our model, 1D conventional DTQW \cite{10.1080/00107151031000110776}, and classical random walk \cite{RN85}, we study the diffusion of the wave function with the initial single-site excitation, i.e., with the initial state $|\varphi_0\rangle = |0\rangle\otimes|H\rangle$. The diffusion distance $M(n)\equiv \sqrt{\sum_{m,p} m^2\big{|}\langle m,p|U^n|\varphi_0\rangle\big{|}^2}$ is defined, which represents the most probable position of the quantum walker after $n$ steps. Figure \ref{fig:fig3} shows the comparison with simulation results of diffusion dynamics of the quantum walker. One sees that the conventional DTQW spreads quadratically faster than the classical random walk in one dimension. Interestingly, the diffusion in the synthetic frequency lattice depends largely on $\mathit{\Gamma}$. For $\mathit{\Gamma}=0.06\pi$, the model is in the weak modulation limit and exhibits the tight-binding feature with only the nearest-neighbor connectivity \cite{RN83,RN73}. The diffusion speed of the walker is slower than that of classical random walks. However, as $\mathit{\Gamma}$ grows to $\pi$ and $3\pi$, the slopes of quasienergy spectrums become sharper and simulation results show much faster diffusion speeds of the walker exceeding that of DTQW. Especially, when $\mathit{\Gamma}=3\pi$, the diffusion speed is accelerated to about 8 times of the speed in DTQW. Moreover, we also find that the wave function distribution during evolutions in synthetic lattice is symmetric, as the translation operator $T$ in our model translates two polarization modes in both directions with the same coupling strength.

The acceleration of the diffusion speed with large $\mathit{\Gamma}$ in our model thanks to the simultaneous translation of the walker over multiple frequency sites, i.e., the diffusion over long distances, but in a coherent way. Moreover, the quantum walk transport in our proposal still holds high efficiency, proven by the calculation of the return probability \cite{mulken2006efficiency,xu2008continuous} (see the supplementary note \uppercase\expandafter{\romannumeral3} for details). With the help of such acceleration, the conversion of the quantum state wave function in the frequency dimension becomes faster. Therefore, such important feature can provide the significant speed-up of various quantum algorithms based on quantum walk.

\subsection{Quantum gate operations}

Due to the coherent multiple long-rang translations [Eq. (\ref{eq10})], our model naturally supports Eq. (\ref{eq11}) and its corresponding band [Eq. (\ref{eq14})], which allows us to implement quantum gate operations in one time step under specific parameters. Basic operations towards quantum operations desire $X$, $Y$, $Z$ gates and also Hadmard ($H$) gate, phase-shift ($R_z$) gate \cite{williams2011quantum}. We consider the qubit with given $k$ in quasi-momentum space $|\phi_1,\phi_2\rangle\otimes|k\rangle$, where $|\phi_1,\phi_2\rangle=\cos(\phi_1/2)|H\rangle+\sin(\phi_1/2)e^{i\phi_2}|V\rangle$, which corresponds to the wave function distribution in the frequency dimension with the Fourier transform. Specific modulations can convert the wave function distribution into the target qubit's distribution on the lattice, equivalent to a quantum gate operation in quasi-momentum space. We take the building of the $X$ gate as an example by setting $\theta=\pi$, $\mathit{\Gamma}\cos(k+\phi_H)=\pi$ and $\mathit{\Gamma}\cos(k+\phi_V)=0$ in Eq. (\ref{eq11}), so the corresponding operator $U_k\equiv\int_{BZ}dk M_U\otimes|k\rangle\langle k|$ has $M_U=
\left(
\begin{array}{cc}
0 & 1 \\
1 & 0 \\
\end{array}\right)\;$
for qubits with definite quasi-momentum $|\phi_1,\phi_2\rangle\otimes|k\rangle$. Similarly, other gates can be obtained by varying parameters [see Table \ref{tab:table1}]. Note here each operation is performed with one roundtrip, thanks to the specific band [Eq. (\ref{eq14})] and large $\mathit{\Gamma}$ we can take, i.e., the step in DTQW required for preparing quantum gates can be greatly reduced (see the supplementary note \uppercase\expandafter{\romannumeral4} for details). In other words, the coherent multiple long-range translations with strong modulation in DTQW can lead to the speed-up of the quantum gate preparation and reduce its time steps into 1 step.

\begin{table}[t]
\caption{\label{tab:table1}Parameters used for realizing quantum gates in quasi-momentum space}
\setlength{\tabcolsep}{10mm}{
\begin{ruledtabular}
\begin{tabular}{c|c|c|c|c}
  & $\theta$ & $\mathit{\Gamma}\cos (k+\phi_H)$ & $\mathit{\Gamma}\cos (k+\phi_V)$ & $M_U$\\
\colrule
$X$ gate & $\pi$ & $\pi$ & 0 & $\left(
\begin{array}{cc}
0 & 1 \\
1 & 0 \\
\end{array}\right)\;$\\

$Y$ gate & \textrm{$\pi$} & $\frac{\pi}{2}$ & $\frac{\pi}{2}$ & $\left(
\begin{array}{cc}
0 & -i \\
i & 0 \\
\end{array}\right)\;$\\

$Z$ gate & 0 & 0 & $\pi$ & $\left(
\begin{array}{cc}
1 & 0 \\
0 & -1 \\
\end{array}\right)\;$\\

$H$ gate & $-\frac{\pi}{2}$ & 0 & $\pi$ & $\left(
\begin{array}{cc}
1 & 1 \\
1 & -1 \\
\end{array}\right)\;$\\

$R_z$ gate & 0 & 0 & $\varphi$ & $\left(
\begin{array}{cc}
1 & 0 \\
0 & e^{i\varphi} \\
\end{array}\right)\;$\\
\end{tabular}
\end{ruledtabular}}
\end{table}

The fidelity of quantum gates are calculated by $F_g=||U_t-U_o||^2_{HS}=Tr[(U_t^{\dagger}-U_o^{\dagger})(U_t-U_o)]$ between the output matrix $U_o$ and the target matrix $U_t$, after simulating the synthetic lattice under the initial excitation with $\Delta=200$, $k=2\pi/3\Omega$, and $\bold{e_s}=(1,0)^T$ or $(0,1)^T$. Simulation results give $F_g=1$, indicating the high fidelity of quantum gate operations in theory.

\begin{figure}[hbpt]
\centering
\includegraphics[width=10cm]{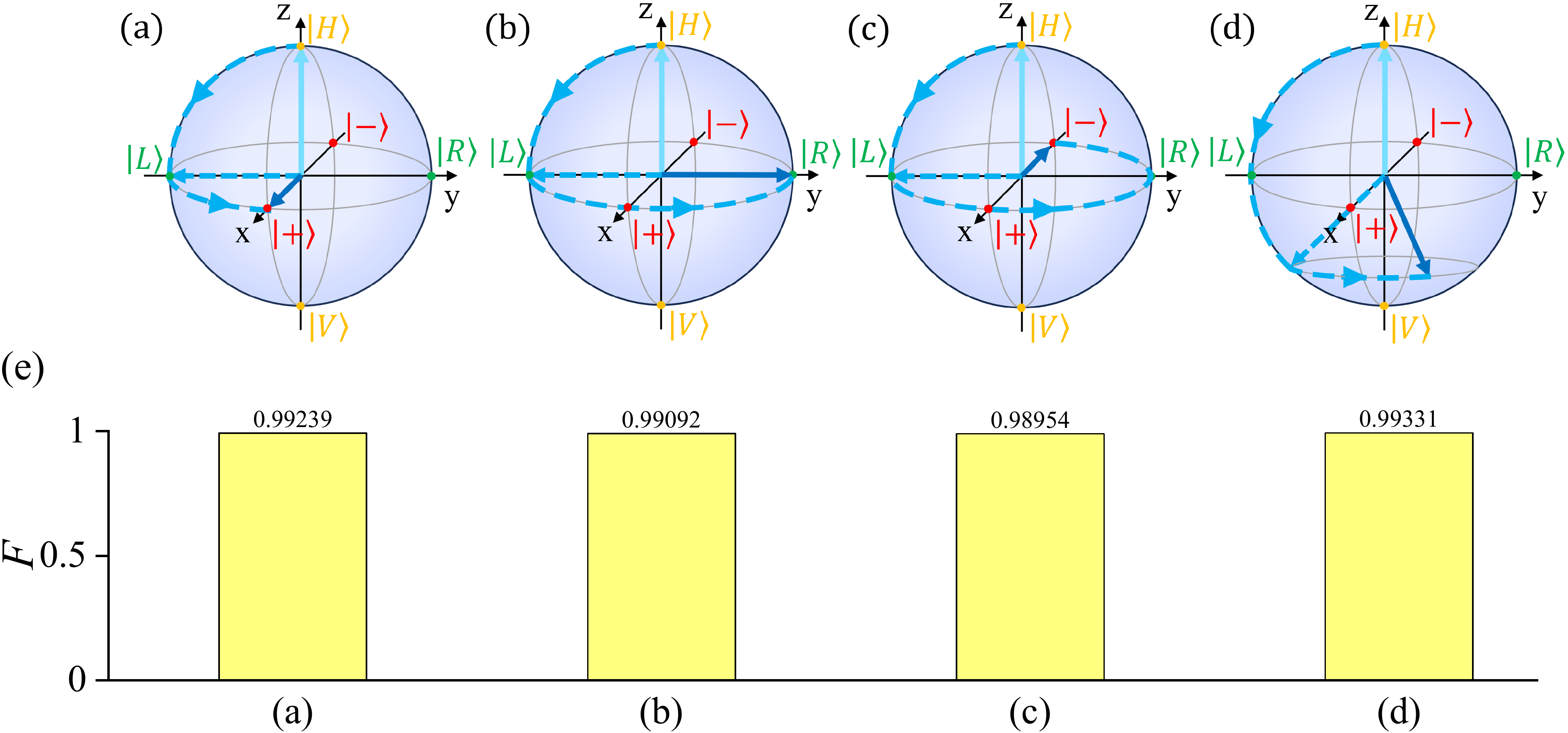}
\caption{\label{fig:fig4} Illustrations of single qubit state preparations using multiple single quantum gates. The targeting single qubit states are (a) $|0.5\pi,0\rangle$, (b) $|0.5\pi,0.5\pi\rangle$, (c) $|0.5\pi,\pi\rangle$ and (d) $|0.75\pi,0.25\pi\rangle$ respectively. (e) The fidelity between these four target single qubit states and their respective output qubit states.}
\end{figure}

Arbitrary single qubit state can then be prepared using the proposed quantum gates. We use two $H$ gates and two $R_z$ gates to illustrate the representative process:
\begin{equation}\label{eq22}
\begin{aligned}
|\phi_1,\phi_2\rangle\otimes|k\rangle=U_P(\phi_2+\frac{\pi}{2})U_HU_P(\phi_1)U_H|H\rangle\otimes|k\rangle.
\end{aligned}
\end{equation}
These four gates can change the state $|H\rangle$ to arbitrary qubit state $|\phi_1,\phi_2\rangle$. The fidelity of quantum states are calculated by $F=|\langle \psi_o|\psi_t\rangle|^2$ between the output qubit $|\psi_o\rangle$ and the target qubit $|\psi_t\rangle$. Four examples are performed with multiple single quantum gate operations illustrated in Fig. \ref{fig:fig4}, where $U_HU_P(\phi_1)U_H$ rotates the initial state $|H\rangle$ by $\phi_1$ about the $x$ axis, and then $U_P(\phi_2+\pi/2)$ rotates it by $\phi_2$ about the $z$ axis to achieve the desired state $|\phi_1,\phi_2\rangle$ adiabatically. As results, high fidelity in simulations are shown in Fig. \ref{fig:fig4}(e). Therefore, our model can realize the sequential action of multiple quantum gates in one resonant ring with high fidelity, which has positive significance for the simplification of quantum circuits operating the single qubit state.

\begin{figure}[hbpt]
\centering
\includegraphics[width=10cm]{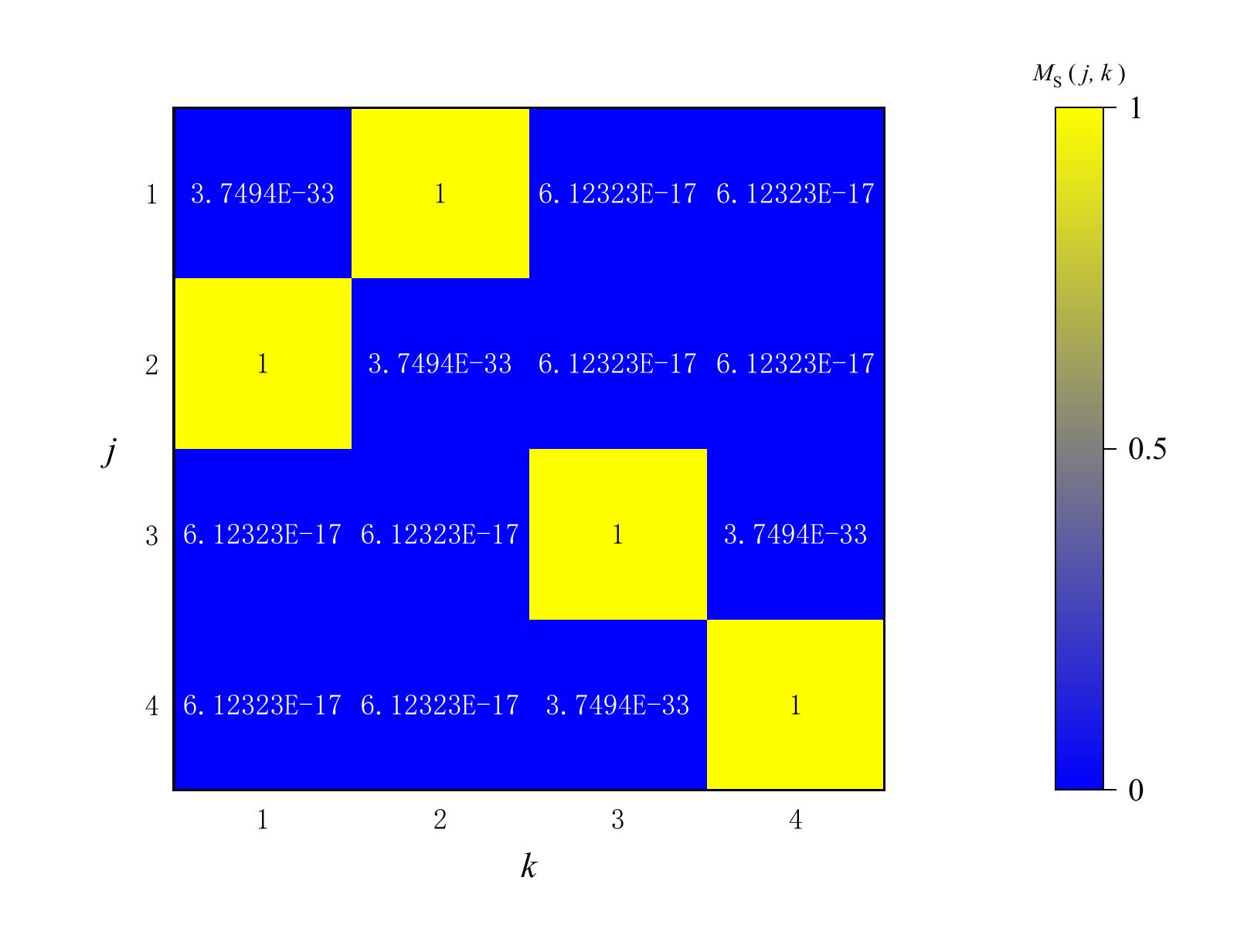}
\caption{\label{fig:fig5}  The reconstructed matrix $M_s$ from simulation results.}
\end{figure}

Our model may be further extended to implement multi-qubit gates by introducing other degrees of freedom \cite{Singh2021}. As an example, we extend our proposal including the position degree of freedom in real space to implement a CNOT gate (see the supplementary note \uppercase\expandafter{\romannumeral5} for details). This model supports the CNOT gate in the form:
\begin{equation}\label{eqS10}
\begin{aligned}
M_{CNOT}=
\left(
\begin{array}{cccc}
1 & 0 & 0 & 0 \\
0 & 1 & 0 & 0 \\
0 & 0 & 0 & 1 \\
0 & 0 & 1 & 0 \\
\end{array}\;\right).
\end{aligned}
\end{equation}
We construct and simulate the following modulation process using two $X$ gates and a CNOT gate:
\begin{equation}\label{eqS11}
\begin{aligned}
M_s=M_XM_{CNOT}M_X=
\left(
\begin{array}{cccc}
0 & 1 & 0 & 0 \\
1 & 0 & 0 & 0 \\
0 & 0 & 1 & 0 \\
0 & 0 & 0 & 1 \\
\end{array}\;\right).
\end{aligned}
\end{equation}
In the simulation, we input different qubits and reconstruct the matrix $M_s$ based on the simulation results as shown in Fig. \ref{fig:fig5}, which is closely matched with the Eq. (\ref{eqS11}). It shows that our model still has high accuracy in the multi-qubit case. Since our proposal is based on the synthetic frequency dimension implemented by a single or limited number of resonant rings, it can be tolerant to the limited spatial scale, especially for applications in the photonic chip regime.

\section{Discussion}

In summary, we propose a method to implement a 1D DTQW system in a synthetic frequency lattice. With strong modulations, we break the weak coupling limit and show multiple long-range couplings can transport the quantum walker largely separated apart over the synthetic lattice but in a coherent way, which induces topological band and faster diffusion speed. The theoretical analysis method of discrete-time quantum walks used in this proposal also fills in the absence of analysis methods for modelling synthetic frequency dimension in the strong modulation limit. Moreover, the band induced by the coherent multiple long-range translations under specific parameters provides the way for constructing single quantum gates in the quasi-momentum space and then the capability for preparing arbitrary single-qubit state as well as multi-qubit gates.

Our proposal is based on geometrically simple but experimental feasible integrated-photonic design \cite{Hu20,RN75,RN77,RN78,hu2022mirror,boes2023lithium,ye2024construction,wang2024chip}, whose requirements for initial state preparations and measurements are experimentally achievable from state-of-art technologies \cite{RN77} (see the supplementary note \uppercase\expandafter{\romannumeral6} for details). Experimentally, high Q-factor is needed to reduce the influence of intrinsic loss on our quantum walk. Recent experiments have proved that it is feasible to build cavities with high Q-factor and perform quantum experiments in the synthetic frequency fimension \cite{RN77,ye2024construction}. Our model with the band under large $\mathit{\Gamma}$ makes quantum gate operations to be implemented with 1 step (roundtrip), thus minimize the impact on the peeformance from loss. The strong-modulation induced quantum walk with intrinsic coherent long-range couplings does not require additional modulation signals at higher frequencies, which potentially provides the important photonic on-chip applications with the quantum algorithm speed-up and the quantum circuit simplification.

\section{acknowledgments}
%\begin{acknowledgments}
The research is supported by the National Key Research and Development Program of China (no. 2023YFA1407200), the National Natural Science Foundation of China (12122407, 12192252, and 12204304), and the China Postdoctoral Science Foundation (2023M742292, GZC20231614).
%\end{acknowledgments}

%

\end{document}